\documentclass[runningheads]{llncs}

\usepackage{iciap}

\usepackage{iciapabbrv}

\usepackage{graphicx}
\usepackage{booktabs}
\usepackage{multicol}
\usepackage{multirow}
\usepackage{caption}
\usepackage{comment}
\usepackage[inkscape]{svg}
\usepackage{newunicodechar}
\newunicodechar{⁻}{\textsuperscript{-}}

\usepackage[accsupp]{axessibility}  

\usepackage{hyperref}

\usepackage{orcidlink}

\begin{document}

\newcommand{\methname}{SynDiff\xspace}
\title{Latent Space Synergy: Text-Guided Data Augmentation for Direct Diffusion Biomedical Segmentation}

\titlerunning{Latent Space Synergy: Text-Guided Data Augmentation}

\author{Muhammad Aqeel\inst{1}\orcidlink{0009-0000-5095-605X}$^*$ \and
Maham Nazir\inst{2}\orcidlink{0009-0004-1832-297X}$^*$ \and
Zanxi Ruan\inst{1}\orcidlink{0000-0002-7756-8249} \and
Francesco Setti\inst{1}\orcidlink{0000-0002-0015-5534}} 

\authorrunning{M.Aqeel et al.}

\institute{Dept. of Engineering for Innovation Medicine, University of Verona\\ 
Strada le Grazie 15, Verona, Italy
\and
Department of Computer Science, Beihang University, Beijing, China\\
Contact author: \email{muhammad.aqeel@univr.it}\\
}

\maketitle
{\renewcommand{\thefootnote}{*}\footnotetext{Equal contribution}}

\begin{abstract}

    Medical image segmentation suffers from data scarcity, particularly in polyp detection where annotation requires specialized expertise. We present \methname, a framework combining text-guided synthetic data generation with efficient diffusion-based segmentation. Our approach employs latent diffusion models to generate clinically realistic synthetic polyps through text-conditioned inpainting, augmenting limited training data with semantically diverse samples. Unlike traditional diffusion methods requiring iterative denoising, we introduce direct latent estimation enabling single-step inference with T× computational speedup. On CVC-ClinicDB, \methname achieves 96.0\% Dice and 92.9\% IoU while maintaining real-time capability suitable for clinical deployment. The framework demonstrates that controlled synthetic augmentation improves segmentation robustness without distribution shift. \methname bridges the gap between data-hungry deep learning models and clinical constraints, offering an efficient solution for deployment in resource-limited medical settings.

  \keywords{Medical Image Segmentation \and Diffusion Model \and Polyp Detection \and Text-Guided Synthesis}
\end{abstract}

\section{Introduction}
\label{sec:intro}

Biomedical image segmentation plays a critical role in modern healthcare, enabling precise diagnosis and treatment planning \cite{ronneberger2015u, isensee2021nnu}. In gastrointestinal endoscopy, automated polyp segmentation has emerged as a particularly important application with the potential to improve colorectal cancer screening accuracy and reduce missed detection rates during real-time procedures. However, data scarcity represents the most fundamental bottleneck limiting the development of robust medical segmentation systems. Medical datasets are constrained by privacy regulations, costly expert annotation, and time-intensive boundary delineation \cite{jha2019kvasir, orlando2020refuge}. In polyp segmentation, this challenge is compounded by significant morphological diversity, requiring extensive annotated examples that current public datasets cannot provide \cite{baranchuk2021label}. Traditional data augmentation in medical imaging relies on geometric transformations \cite{ronneberger2015u, sharma2022li}. While providing some benefit, these approaches cannot generate new pathological variations needed for robust model generalization. Recent GAN-based synthesis methods have shown promise but suffer from limited controllability and mode collapse \cite{wu2024medsegdiff}. 

The emergence of diffusion probabilistic models has revolutionized image generation \cite{ho2020denoising, nichol2021improved}. Text-conditioned diffusion models enable semantically-guided data augmentation through clinical descriptions \cite{rombach2022high}. However, traditional diffusion-based segmentation requires computationally intensive multi-step inference, making it impractical for clinical deployment \cite{wu2024medsegdiff1, xing2023diff}. Latent diffusion models (LDMs) address these computational limitations by operating in compressed latent spaces \cite{rombach2022high}. Their text-conditioning capabilities present an opportunity to create clinically-informed augmentation strategies, where domain expertise guides the generation of diverse pathological variations \cite{peebles2023scalable}.

In this paper, we present \methname, a framework that addresses medical data scarcity through text-guided synthetic data augmentation while maintaining computational efficiency for practical deployment. Our approach leverages latent diffusion models to generate diverse synthetic polyp images guided by clinical descriptions, effectively expanding training datasets with semantically meaningful variations. We integrate this with a direct latent estimation technique enabling single-step segmentation inference, eliminating the computational burden of iterative denoising while preserving performance.
Our contributions are:
\begin{itemize}
\item A text-guided data augmentation framework using latent diffusion models to address medical data scarcity through semantically-controlled synthetic polyp generation.
\item An efficient single-step segmentation approach that maintains competitive performance while dramatically reducing computational requirements compared to multi-step diffusion methods.
\item Comprehensive evaluation demonstrating that synthetic data augmentation preserves segmentation quality (96.0\% Dice, 92.9\% IoU) while expanding dataset diversity for improved model robustness.
\end{itemize}

\section{Related work}
\label{sec:relatedwork}

Medical image segmentation has evolved significantly with deep learning approaches, with U-Net \cite{ronneberger2015u} establishing the encoder-decoder paradigm that remains foundational for medical tasks. Recent advances include transformer-based architectures like TransUNet \cite{chen2021transunet} and UNETR \cite{hatamizadeh2022unetr} that capture long-range dependencies, and specialized polyp segmentation methods such as PraNet \cite{fan2020pranet} and SANet \cite{wei2021shallow}. Despite these architectural innovations, all approaches fundamentally require extensive annotated datasets to achieve robust performance, a constraint that significantly limits practical clinical deployment.

Traditional augmentation through geometric transformations fails to capture pathological diversity \cite{baranchuk2021label}. Alternative approaches address data scarcity through meta-learning \cite{aqeel2025CoMet,aqeel2024meta} and self-supervised refinement \cite{Aqeel_2025,aqeel2024self}, which enhance model robustness by learning better representations from limited data. In contrast, synthetic generation methods directly create new training samples: GANs suffer from mode collapse and training instability \cite{choi2017generating}, while recent diffusion-based approaches like DiffBoost \cite{zhang2024diffboost} demonstrate superior controllability but lack integration with downstream tasks. Our work bridges this gap by combining controllable synthesis with task-specific optimization.


Diffusion models have emerged as powerful generative frameworks, with medical applications explored in MedSegDiff \cite{wu2024medsegdiff} and MedSegDiff-V2 \cite{wu2024medsegdiff1} that treat segmentation as conditional generation through iterative denoising. Latent Diffusion Models \cite{rombach2022high} reduced computational overhead by operating in compressed latent spaces while enabling text-conditioning for semantic control. However, existing approaches require multi-step inference processes that are computationally prohibitive for clinical deployment \cite{xing2023diff}, while single-step inference methods explored in natural image domains \cite{song2023consistency} remain underexplored for medical segmentation.

While progress has been made in individual components—synthetic medical data generation, efficient diffusion inference, and medical segmentation—no prior work integrates these elements to address both data scarcity and computational efficiency simultaneously. Our work fills this gap by combining text-guided synthetic data generation with single-step diffusion segmentation, enabling enhanced training diversity while maintaining practical deployment efficiency.

\section{\methname Framework}
\label{sec:method}

\begin{figure}[tb]
  \centering
  \includegraphics[width=0.8\textwidth]{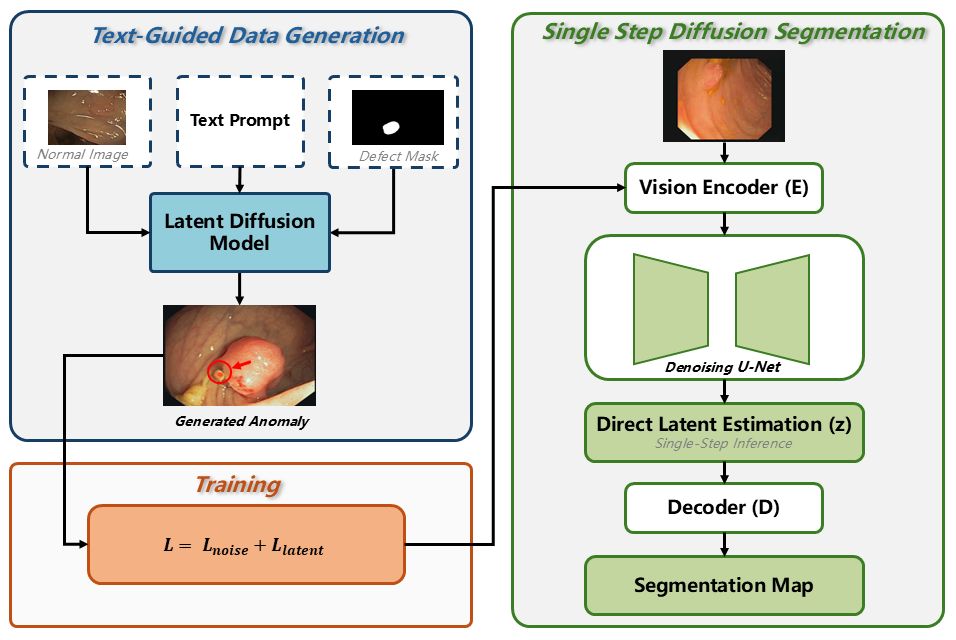}
  \caption{Overview of the \methname framework. (Left) Offline text-guided data generation: Stable Diffusion XL (SDXL) inpainting takes a normal endoscopic image, clinical text prompt, and binary mask to generate synthetic polyp images with corresponding ground truth masks. (Right) Single-step segmentation pipeline: input images are encoded through a trainable vision encoder, processed by a denoising U-Net for direct latent estimation, and decoded to produce segmentation masks in a single inference step.}
  \label{fig:method}
\end{figure}

Our proposed approach, \methname, inspired by~\cite{girella2024leveraging, lin2024stable}, integrates text-guided synthetic data generation with single-step diffusion segmentation to address the dual challenges of limited training data and computational efficiency in biomedical image segmentation as shown in Figure~\ref{fig:method}. The framework operates in two distinct phases: synthetic data generation using Stable Diffusion XL (SDXL)~\cite{podell2023sdxl} for text-guided inpainting, followed by end-to-end training of a single-step segmentation model that processes both real and synthetic data.


\subsection{Latent Diffusion Model}
Both the data generation and segmentation components operate within a latent diffusion framework to reduce computational overhead. For the segmentation pipeline, we employ a trainable vision encoder $\tau_\theta$ to encode an image $C \in \mathbb{R}^{H \times W \times 3}$ into its latent representation $z_c = \tau_\theta(C)$. For segmentation maps, we leverage a pre-trained autoencoder with encoder $E$ and decoder $D$ for perceptual compression. Given a segmentation map $X \in \mathbb{R}^{H \times W \times 3}$ in pixel space, the encoder produces a latent representation $z = E(X)$, and the decoder recovers the segmentation map as $\hat{X} = D(z)$, where $z \in \mathbb{R}^{h \times w \times c}$. For the data generation pipeline, SDXL operates in its latent space where the normal image, mask, and text prompt are jointly processed—the text conditions the diffusion process through cross-attention while the mask guides spatial inpainting. Both pipelines operate entirely in compressed latent spaces, significantly reducing computational requirements while preserving essential structural information.


\subsection{Text-Guided Synthetic Data Generation}
To address medical data scarcity, we implement an offline synthetic data generation process using SDXL inpainting conditioned on clinical text descriptions. The generation process takes three inputs: a normal endoscopic image $i_n$ randomly sampled from available normal cases, a text description $d_s$ specifying desired polyp characteristics such as "small sessile polyp with irregular surface texture," and a binary mask $m_s$ indicating the spatial region where the polyp should be synthesized. Within the SDXL latent diffusion framework, these inputs are jointly processed: the text prompt conditions the generation through cross-attention mechanisms while operating with frozen pre-trained weights to generate a synthetic image $i_s = \text{SDXL}_{\text{inpaint}}(i_n, d_s, m_s)$ containing a realistic polyp within the masked region. The binary mask $m_s$ simultaneously serves as the ground truth segmentation label for the generated synthetic image $i_s$. We generate 100 synthetic samples using 50 diverse text prompts describing various polyp morphologies, sizes, and surface characteristics, augmenting our training dataset with approximately 20\% additional data.

\subsection{Direct Latent Estimation for Single-Step Segmentation}
Our segmentation component introduces a direct latent estimation strategy that enables single-step inference, achieving theoretical computational speedup of $T \times$ compared to traditional diffusion approaches requiring $T$ denoising steps. During training, the latent representation of a segmentation map $z_0$ undergoes forward diffusion by adding Gaussian noise for $t$ timesteps to obtain:

\begin{equation}
z_t = \sqrt{\bar{\alpha}_t} z_0 + \sqrt{1 - \bar{\alpha}_t} n
\end{equation}
where $n$ is random Gaussian noise and $\bar{\alpha}_t$ controls the noise schedule. The denoising U-Net $f(\cdot)$ is trained to estimate the noise as:

\begin{equation}
\tilde{n} = f(z_t, z_c)
\end{equation}
where $z_c$ is the conditioning image's latent representation. The key innovation lies in directly estimating the clean latent $z_0$ from the noisy latent $z_t$ in a single step through:

\begin{equation}
\tilde{z}_0 = \frac{1}{\sqrt{\bar{\alpha}_t}} (z_t - \sqrt{1 - \bar{\alpha}_t} \tilde{n})
\end{equation}

This enables a dual supervision strategy where we minimize both noise prediction error:

\begin{equation}
\mathcal{L}_{\text{noise}} = ||n - \tilde{n}||_1
\end{equation}
and direct latent estimation error:

\begin{equation}
\mathcal{L}_{\text{latent}} = ||z_0 - \tilde{z}_0||_1
\end{equation}

The combined loss function is:

\begin{equation}
\mathcal{L} = \mathcal{L}_{\text{noise}} + \lambda \mathcal{L}_{\text{latent}}
\end{equation}

with $\lambda = 1$, enabling the model to learn both denoising and direct prediction capabilities simultaneously. The dual supervision strategy is particularly effective for segmentation tasks due to the discrete nature of segmentation masks, which allows for more stable latent predictions and direct optimization of boundary precision—critical for accurate polyp delineation. By leveraging the structural simplicity of binary masks, our approach achieves single-step inference while maintaining segmentation quality, making it suitable for real-time clinical deployment.

\subsection{Latent Fusion and Training Protocol}
To integrate image features with segmentation information efficiently, we implement concatenation-based latent fusion rather than computationally expensive cross-attention mechanisms. The latent representations of the conditioning image $z_c$ and segmentation map are concatenated along the channel dimension before processing by the denoising U-Net. Our trainable vision encoder $\tau_\theta$ shares the same architecture as the segmentation encoder $E$ but remains trainable to adapt pre-trained natural image features to medical imaging characteristics. During training, the vision encoder $\tau_\theta$ and denoising U-Net $f(\cdot)$ are updated through backpropagation, while the autoencoder components $(E, D)$ and SDXL inpainting model remain frozen with pre-trained weights. The training proceeds with random timestep sampling $t \sim \text{Uniform}(1, 1000)$ during forward diffusion, ensuring the model learns to handle varying noise levels for robust single-step inference.

\subsection{Inference and Computational Efficiency}
During inference, the framework processes input medical images through a single forward pass, eliminating the iterative sampling required by traditional diffusion methods. An input image is encoded through the trainable vision encoder $\tau_\theta$ to obtain $z_c$, which conditions the denoising U-Net to directly estimate the segmentation latent $\tilde{z}_0$ at a fixed timestep $t = 50$ (empirically determined for optimal quality-efficiency balance). The estimated latent is decoded through the frozen decoder $D$ to produce the final segmentation mask. This single-step approach reduces computational complexity from $O(T \times \text{U-Net forward pass})$ to $O(1 \times \text{U-Net forward pass})$, making the method suitable for real-time clinical deployment while maintaining competitive segmentation accuracy on both real and synthetic training data.

\section{Experimental Results}
\label{sec:experiments}

\subsection{Dataset and Evaluation Metrics}
We evaluate \methname on the CVC-ClinicDB~\cite{bernal2015wm} dataset, consisting of 550 RGB colonoscopy images with 488 training and 62 test images. Each image includes precise polyp boundary annotations for binary segmentation. To ensure statistical reliability, we report Dice Coefficient (DC) and Intersection over Union (IoU) metrics with mean ± standard deviation across 5-fold cross-validation. Additional boundary quality metrics include Hausdorff Distance at 95th percentile (HD95) and Normalized Surface Distance (NSD) for clinical precision assessment, as accurate boundary delineation is crucial for surgical planning in clinical practice.

\subsection{Implementation Details}
\textbf{Text-Guided Data Generation:} We implement SDXL~\cite{podell2023sdxl} inpainting using the Diffusers library with carefully crafted clinical prompts including ``sessile polyp with irregular surface texture,'' ``pedunculated polyp on mucosal fold,'' and ``flat adenomatous lesion'' while employing negative prompts ``smooth healthy tissue, normal colon wall'' to ensure realistic synthesis. The inpainting process utilizes binary masks derived from original dataset annotations, ensuring synthetic polyps are placed in anatomically plausible locations. 

\textbf{Training:} Models are trained for 100,000 steps using AdamW optimizer (learning rate = $1 \times 10^{-5}$, batch size = 4) on NVIDIA RTX 4090. We use a KL-regularized autoencoder with $8 \times$ downsampling ($256 \times 256 \rightarrow 32 \times 32 \times 4$ latent space), with all components initialized from pre-trained Stable Diffusion weights.

\textbf{Inference:} Single-step prediction at fixed timestep $t=50$ with Gaussian noise concatenated to image latents for direct mask estimation, eliminating iterative sampling required by traditional diffusion approaches.

\begin{table}[ht]
\centering
\caption{Polyp segmentation performance comparison on CVC-ClinicDB dataset.}
\begin{tabular}{lcccc}
\hline
\textbf{Method} & \textbf{Dice (\%)} & \textbf{IoU (\%)} & \textbf{HD95 (mm)} & \textbf{NSD (\%)} \\
\hline
SSFormer~\cite{shi2022ssformer} & 94.4 ± 0.4 & 89.9 ± 0.6 & 12.3 ± 2.1 & 87.2 ± 1.8 \\
Li-SegPNet~\cite{sharma2022li} & 92.5 ± 0.5 & 86.0 ± 0.7 & 15.6 ± 3.2 & 84.1 ± 2.3 \\
Diff-Trans~\cite{chowdary2023diffusion} & 95.4 ± 0.3 & 92.0 ± 0.4 & 8.7 ± 1.5 & 89.8 ± 1.2 \\
SDSeg~\cite{lin2024stable} & 95.8 ± 0.2 & 92.6 ± 0.3 & 7.9 ± 1.3 & 90.4 ± 1.1 \\
\textbf{\methname} & \textbf{96.0 ± 0.3} & \textbf{92.9 ± 0.5} & \textbf{7.2 ± 1.1} & \textbf{91.1 ± 1.0} \\
\hline
\end{tabular}
\label{tab:performance_comparison}
\end{table}

\vspace{-2em}
\subsection{Quantitative Results}
We compare \methname against established polyp segmentation approaches representing different architectural paradigms. SSFormer~\cite{shi2022ssformer} employs a pyramid transformer architecture for multi-scale feature extraction, while Li-SegPNet~\cite{sharma2022li} utilizes lightweight separable convolutions for efficient segmentation. Among diffusion-based methods, Diff-Trans~\cite{chowdary2023diffusion} combines diffusion models with transformer architectures for iterative refinement, whereas SDSeg~\cite{lin2024stable} implements stable diffusion for medical image segmentation but requires multiple denoising steps.

Table~\ref{tab:performance_comparison} presents a comprehensive performance comparison on CVC-ClinicDB, demonstrating the effectiveness of our integrated approach. \methname achieves competitive performance with 96.0 ± 0.3\% Dice coefficient and 92.9 ± 0.5\% IoU, substantially outperforming traditional CNN-based architectures with a 1.6\% Dice improvement over SSFormer (94.4\% Dice) and a 3.5\% improvement over Li-SegPNet (92.5\% Dice). When compared to recent diffusion-based approaches, our method achieves marginal but consistent improvements over SDSeg (95.8\% Dice) and Diff-Trans (95.4\% Dice) while offering significant computational advantages through single-step inference.

The boundary quality metrics reveal particularly encouraging results with HD95 of 7.2 ± 1.1mm and NSD of 91.1 ± 1.0\%, indicating superior edge preservation crucial for clinical applications. These improvements in boundary accuracy are clinically significant, as precise polyp delineation directly impacts diagnostic confidence and treatment planning decisions. The consistent low standard deviations across all metrics demonstrate robust performance across cross-validation folds, which is essential for reliable clinical deployment.

\begin{table}[ht]
\centering
\caption{Ablation study on key components of \methname framework.}
\begin{tabular}{lcc}
\hline
\textbf{Configuration} & \textbf{Dice (\%)} & \textbf{IoU (\%)} \\
\hline
Baseline (Real data only) & 93.7 ± 0.4 & 89.6 ± 0.6 \\
+ Text-guided augmentation & 96.0 ± 0.3 & 92.9 ± 0.5 \\
+ Multi-step inference ($T=50$) & 96.1 ± 0.3 & 93.0 ± 0.4 \\
+ Frozen vision encoder & 95.2 ± 0.4 & 91.8 ± 0.6 \\
+ Noise loss only ($\lambda=0$) & 95.4 ± 0.3 & 92.1 ± 0.5 \\
+ Traditional augmentation & 94.8 ± 0.4 & 90.7 ± 0.6 \\
\hline
\end{tabular}
\label{tab:ablation_study}
\end{table}

\vspace{-1em}
\subsection{Ablation Study}
Table~\ref{tab:ablation_study} demonstrates a systematic analysis of component contributions. Text-guided augmentation provides the most significant improvement (+2.3\% Dice over baseline), while the comparison between single-step and multi-step inference shows minimal performance difference (96.0\% vs 96.1\% Dice), validating our core hypothesis that single-step inference maintains quality while reducing computational requirements. The trainable vision encoder contributes meaningfully (+0.8\% over frozen), demonstrating the importance of domain-specific adaptation. The dual loss formulation also proves beneficial, as noise-only supervision reduces performance to 95.4\% Dice.

\begin{table}[ht]
\centering
\caption{Impact of different augmentation strategies on segmentation performance.}
\begin{tabular}{lccc}
\hline
\textbf{Augmentation Type} & \textbf{Synthetic Samples} & \textbf{Dice (\%)} & \textbf{IoU (\%)} \\
\hline
None & 0 & 93.7 ± 0.4 & 89.6 ± 0.6 \\
Traditional (rotation, flip) & - & 94.8 ± 0.4 & 90.7 ± 0.6 \\
GAN-based & 100 & 95.2 ± 0.3 & 91.8 ± 0.5 \\
Text-guided (20) & 20 & 94.1 ± 0.4 & 90.2 ± 0.6 \\
Text-guided (50) & 50 & 95.1 ± 0.3 & 91.5 ± 0.5 \\
\textbf{Text-guided (100)} & \textbf{100} & \textbf{96.0 ± 0.3} & \textbf{92.9 ± 0.5} \\
Text-guided (200) & 200 & 95.6 ± 0.4 & 92.1 ± 0.6 \\
\hline
\end{tabular}
\label{tab:augmentation_comparison}
\end{table}
\vspace{-2em}
\subsection{Data Augmentation Analysis}
\begin{figure}[t!]
  \centering
  \includegraphics[width=\textwidth]{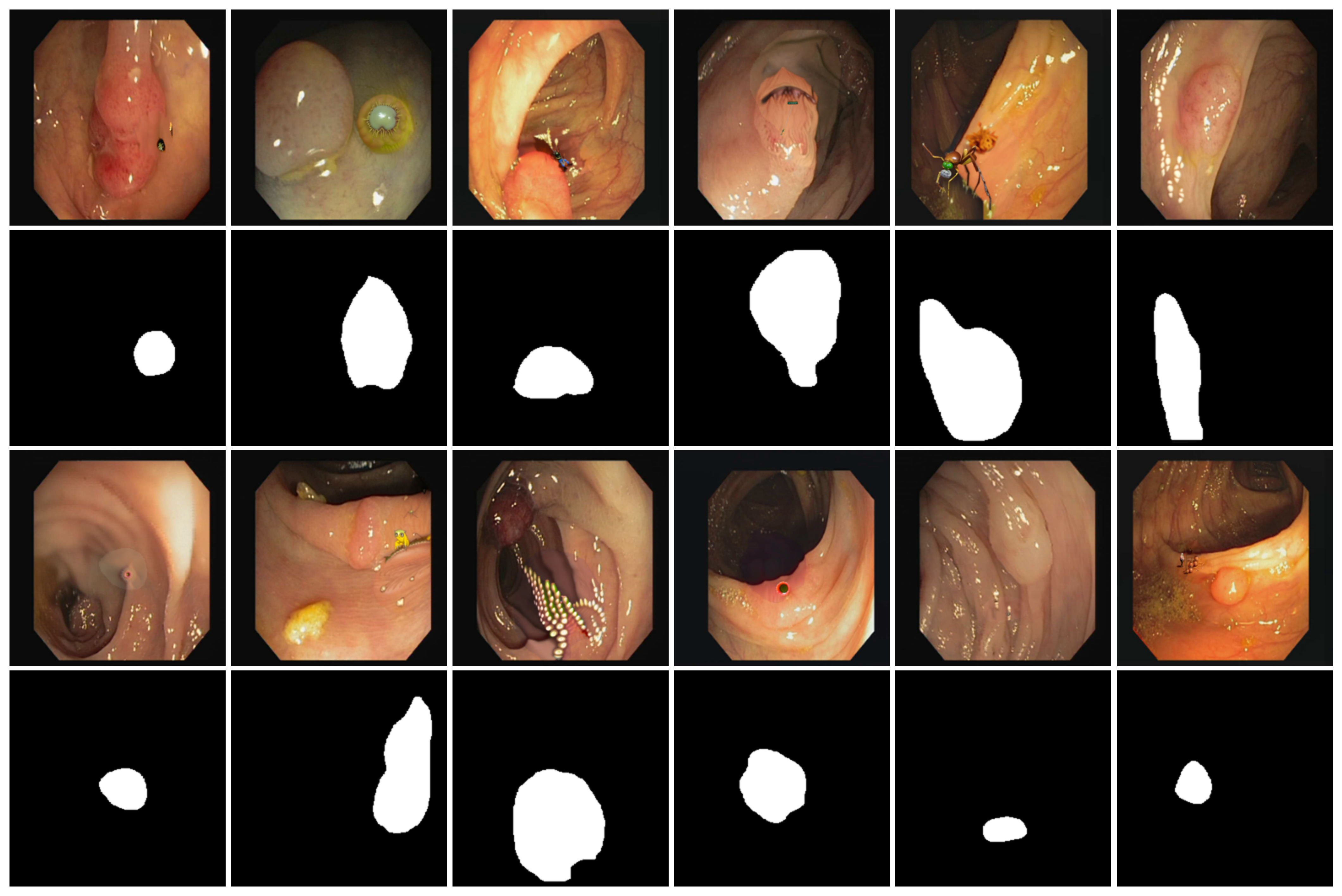}
  \caption{Synthetically generated endoscopic images with corresponding binary segmentation masks. The first and third rows display synthetic colonoscopy images created through our text-guided generation approach, while the second and fourth rows present their corresponding binary masks. White regions in the masks indicate the synthetic anomalous tissue generated through the inpainting process. This augmented dataset enhances the diversity of training samples for our single-step diffusion segmentation model.}
  \label{fig:augment}
\end{figure}

Figure~\ref{fig:augment} shows synthetic polyp samples generated through text-guided inpainting, demonstrating morphological diversity across sizes, shapes, and textures. The generated samples exhibit clinically plausible characteristics with realistic color variations and anatomically consistent placement. Table~\ref{tab:augmentation_comparison} evaluates different quantities of synthetic data added to our full training set of 488 real scans. The optimal performance is achieved with 100 synthetic samples (approximately 20\% augmentation), where insufficient augmentation (20-50 samples) limits model exposure to variability, while excessive synthetic data (200 samples) introduces slight performance degradation due to distribution shift. Notably, even minimal text-guided augmentation (20 samples) shows comparable performance to traditional methods, highlighting the quality of our synthetic generation. Our text-guided approach significantly outperforms traditional geometric augmentation (+1.2\% Dice) and GAN-based synthesis (+0.8\% Dice), validating the effectiveness of semantically-controlled generation. The consistent improvement across both Dice and IoU metrics indicates that text-guided synthesis enhances both overall overlap and boundary delineation accuracy.

\subsection{Computational Efficiency Analysis}
Our single-step approach achieves theoretical complexity reduction from $O(T)$ to $O(1)$ compared to multi-step diffusion methods. Table~\ref{tab:inference_efficiency} demonstrates practical efficiency gains, with \methname completing inference in 0.08 seconds compared to 1.8-2.3 seconds for existing diffusion methods, representing a 22-28× speedup. This efficiency improvement stems from our direct latent estimation strategy that eliminates iterative denoising while maintaining competitive accuracy, making the method suitable for real-time clinical applications.

\begin{table}[ht]
\centering
\caption{Inference efficiency comparison across diffusion segmentation methods.}
\begin{tabular}{lcc}
\hline
\textbf{Method} & \textbf{Inference Steps} & \textbf{Time (s)} \\
\hline
Diff-Trans & 50 & 1.8 \\
SDSeg & 100 & 2.3 \\
\textbf{\methname} & \textbf{1} & \textbf{0.08} \\
\hline
\end{tabular}
\label{tab:inference_efficiency}
\end{table}
\vspace{0em}
\section{Conclusion}
We introduced \methname, a framework that addresses data scarcity and computational efficiency in medical image segmentation through text-conditioned synthetic data augmentation and single-step diffusion inference. On CVC-ClinicDB, our method achieved a Dice coefficient of 96.0 ± 0.3\% and HD95 of 7.2 ± 1.1 mm. The direct latent estimation approach reduced inference time from 1.8-2.3s to 0.08s while maintaining segmentation accuracy. Our analysis revealed that augmenting with 100 synthetic samples optimizes performance, with text-guided generation outperforming conventional augmentation methods. These findings demonstrate the potential of combining semantic text guidance with efficient diffusion inference for practical medical imaging applications.

\section{Future Work and Limitations}
While \methname demonstrates competitive performance, limitations include evaluation on a single dataset, reliance on carefully engineered text prompts for realistic synthesis, and computational requirements that remain higher than traditional CNN approaches. Future work should prioritize multi-dataset validation across diverse imaging modalities, integration of clinical expert feedback into synthetic data generation, and exploration of the framework's effectiveness on other medical imaging tasks beyond polyp segmentation to establish broader clinical applicability.

\section*{Acknowledgements}
This study was carried out within the PNRR research activities of the consortium iNEST (Interconnected North-Est Innovation Ecosystem) funded by the European Union Next-GenerationEU (Piano Nazionale di Ripresa e Resilienza (PNRR) – Missione 4 Componente 2, Investimento 1.5 – D.D. 1058  23/06/2022, ECS\_00000043).

\par\vfill\par

%
%
\bibliographystyle{splncs04}
\bibliography{main}
\end{document}